\documentclass[pre,twocolumn,showpacs,amsmath,amssymb]{revtex4}
\usepackage{epsfig}
\newcommand{\etal}{\emph{et al.}} \newcommand{\ud}{\mathrm{d}}
\begin{document}
%--------------------------------------------------------------------
% ------------ Title & affiliations ------------
%--------------------------------------------------------------------
\title{Highly sensitive refractometer with photonic crystal fiber
long-period grating}
%------------------------------------------------------------------
\author{Lars Rindorf and Ole Bang} \affiliation{COM{$\bullet$}DTU,
Department of Communications, Optics and Materials, Technical
University of Denmark, DK-2800 Kgs. Lyngby, Denmark}
%--------------------------------------------------------------------
% ------------ Abstract ------------
%--------------------------------------------------------------------
\begin{abstract} We present highly sensitive refractometers based on
a long-period grating in a large mode area PCF. The maximum
sensitivity is 1500 nm/RIU at a refractive index of 1.33, the
highest reported for any fiber grating. The minimal detectable index
change is $2\times 10^{-5}$. The high sensitivity is obtained by
infiltrating the sample into the holes of the photonic crystal fiber
to give a strong interaction between the sample and the probing
field. \end{abstract}
 \maketitle
%--------------------------------------------------------------------
% ------------ Introduction ------------
%--------------------------------------------------------------------
Optical fiber sensors are attracting increasing interest.  Fiber
grating sensors are being used for a variety of purposes including
temperature, strain, and refractive index sensing.  The sensors
possess high sensitivity as well as low susceptibility to
interferences. In long-period fiber gratings (LPGs) a core mode is
coupled resonantly to a cladding mode. In standard optical fibers
 the cladding mode probes the surroundings of the fiber and in this
way the resonance wavelength may be shifted. The shift in resonance
wavelength is used as the indicator of the refractometer
\cite{bhatia1996}. At a refractive index of 1.33, typical for
aqueous environments, the typical sensitivity of an LPG in a
standard telecom fiber is typically 50 nm/RIU \cite{zhu2004b}.

Long-period gratings can also be realized in photonic crystal fibers
(PCFs) \cite{eggleton:1999}. PCFs have an array of air holes running
along the fiber axis, which confine the light to the core. The
propagating wave inside the PCF has a particular strong evanescent
wave compared with a standard optical fiber due to a much closer
proximity of the electromagnetic wave and the holes than the outside
of the cladding. PCFs are characterized by their hole diameter, $d$,
and the pitch of the structure, $\Lambda$. Fini \cite{fini2004} has
shown that the probing of the holes is strong when the structure has
a large air filling fraction, and the feature size is comparable to
the wavelength, i.e. for small pitch and for large air holes. The
probing also increases when the contrast between the refractive
index of silica and the refractive index of the holes is small. Phan
Huy \etal\ \cite{phanhuy2007} has recently studied the sensitivity
of a PCF with a Bragg grating to refractive index. By fabricating a
PCF with a very small core, and thus large evanescent field, $f_u$,
the sensitivity was increased by two orders of magnitude with
respect to a large core design.
%--------------------------------------------------------------------

% ------------ Motivation ------------
%--------------------------------------------------------------------
In this Letter we show that the refractive index sensitivity for a
photonic crystal fiber long-period grating is almost two orders of
magnitude larger than the sensitivity for long-period grating for a
standard optical fibers. The sensitivity is  enhanced by a factor of
three by choosing longer wavelengths. Finally, guidelines to
PCF-LPGs with even surpassing sensitivity are discussed.
%--------------------------------------------------------------------

% ------------ Background ------------
%--------------------------------------------------------------------
%---------------------------------------------------------------------------
%\begin{figure}[b!] \begin{center} \includegraphics[angle = 0, width
%=0.4\textwidth]{LMA10_SEM.eps} \end{center}
%\caption{Scanning electron microscope pictures of the LMA10 PCF
%\cite{cf} used. The parameters are $d/\Lambda = 0.478$, $\Lambda =
%7.12\,\mu$m.}\label{fig:sem} \end{figure}
%---------------------------------------------------------------------------
%---------------------------------------------------------------------------
 \begin{figure}[b!]
\begin{center} \includegraphics[angle = 0, width
=0.5\textwidth]{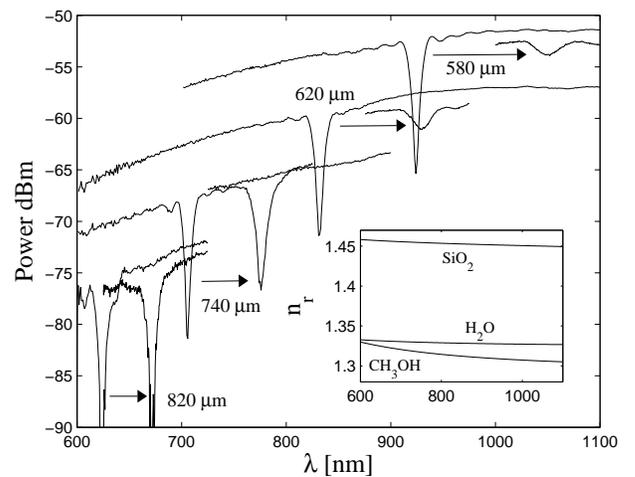} \end{center} \caption{The spectra
of the PCF-LPG filled with air and methanol. Each pair is offset by
5 dBm for clarity. Inset: the refractive index of silica glass,
water and methanol.}\label{fig:spec} \end{figure}
%---------------------------------------------------------------------------

%---------------------------------------------------------------------------
In an LPG the incident core mode can be coupled resonantly to a
cladding mode by a period perturbation that equals the beat length
between the two modes. According to coupled mode theory
\cite{yariv1973} the resonance condition is $\lambda_{\rm r} =
(n_{\rm co}(\lambda)- n_{\rm cl}(\lambda))\Lambda_{\rm G}$. The
effective indices also depend on wavelength, and this must be taken
into considerations. The PCF used was a large mode area PCF with a
mode diameter of 10 $\mu$m. The structure parameters were $\Lambda =
7.12 \mu$m and $d/\Lambda \simeq 0.478$ making
the PCF almost single-mode. The long-period gratings were inscribed
with the CO$_2$ laser method. We used a Synrad Fenix CO$_2$ laser
with a maximum output power of 75 W which is set to 3 \%. There was
no collapsing of the holes, which has also been observed by others
\cite{zhu2003}. In our experimental setup the PCF is mounted in a
stage on top of two linear stages. The stages move the fiber in and
out of the CO$_2$ laser beam, which is kept fixed. The stages are
controlled by a Labview program, and the inscription is fully
automated. The inscription progress is monitored in situ. The setup
gives a high degree of freedom by the choice of the number of
grating periods and the grating period, $\Lambda_{\rm G}$.
%---------------------------------------------------------------------------

%---------------------------------------------------------------------------
Four PCF-LPGs are fabricated with grating periods 820, 740, 620, 580
$\mu$m. The number of periods is 60, the largest number reported to
date for a CO$_2$ laser inscribed PCF-LPG, making the total lengths
49, 44, 37, 35 mm, respectively. The spectra and the resonance
wavelength are presented in Fig.~\ref{fig:spec} and Table
\ref{table}. The spectra show only a single resonance in contrast to
LPGs in standard optical fibers which have multiple resonances. The
PCF-LPGs were 30-40 cm in length. One end was directly connected to
an ANDO AQ-6315A Optical Spectrum Analyzer and the other end was
connected to a broadband halogen light source (Ocean Optics
HL-2000).
%---------------------------------------------------------------------------

%---------------------------------------------------------------------------
Methanol has a refractive index close to that of  water.  Their
refractive indices are displayed in Fig.~\ref{fig:spec} along with
the refractive index of silica calculated from a Sellmeier
expression. The refractive indices of water and methanol are found
from empirical Cauchy expressions \cite{el-kashef}, $n = A +
B/\lambda^2$, with the fitted parameters (400-800 nm): methanol $A =
1.29461 \pm 1\times10^{-5}, B= 12706.1\pm 0.1$nm$^2$, water $A =
1.3242 \pm 1\times 10^{-5}, B = 3063.799 \pm 0.031$nm$^2$. The
thermo-optic coefficient of methanol is dependent on the wavelength
but can be found using the equation of Murphy and Alpert
\cite{el-kashef}.
%---------------------------------------------------------------------------

%---------------------------------------------------------------------------
The methanol is infiltrated into the PCFs by immersing one end
inside a pressure chamber with the other end outside. A 200 kPa
overhead was applied for an hour, after which no bubbles were
observed at the exit facet. The methanol is degassed 30 min in
vacuum prior to infiltration to avoid the formation of air bubbles
inside the PCF. The resulting spectra are seen in
Fig.~\ref{fig:spec}. The magnitude of the shifts clearly increases
with the resonance wavelength. The resonance wavelengths are
obtained by interpolating the transmission (on a linear scale)
around the resonance dip with a second order polynomial. Large red
shifts of 48, 72, 97, and 127 nm in the resonance wavelength are
seen. With such large wavelength shift linearity can not be
expected. The sensitivity itself increases with increasing
refractive index. To obtain the correct sensitivities at a
refractive index of the liquid we tune the refractive index by
temperature through the thermo-optic coefficient of the liquid. The
temperature response of the PCF-LPG with air in the holes is
negligible, a mere $\sim$6 pm/$^\circ$C was  measured.
%---------------------------------------------------------------------------

\begin{figure}[b!] \begin{center} \includegraphics[angle = 0, width
=0.5\textwidth]{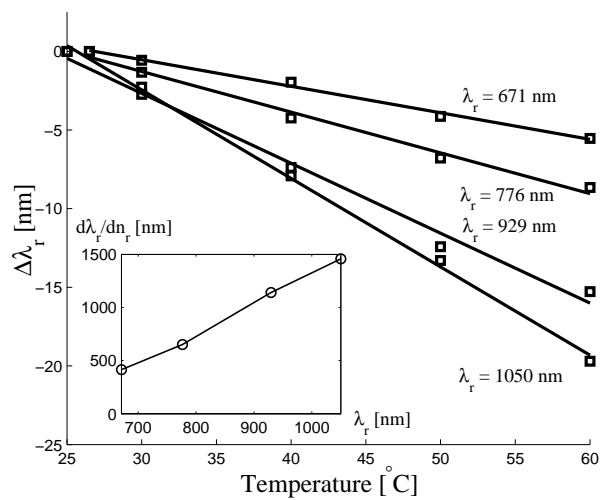} \end{center} \caption{Shifts for
different resonance wavelengths for methanol filled LPGs. Inset: the
refractive index coefficient for the different resonant wavelengths.
}\label{fig:sens} \end{figure}

%---------------------------------------------------------------------------
The PCF-LPGs were mounted onto a heater stage with temperature
control (MC60 \& TH60, Linkam Scientific Instruments). The
temperature was increased in steps up to 60$^\circ$C. The refractive
index of the methanol decreases with temperature, since the
thermo-optic coefficient is negative  (Table \ref{table}) thereby
blueshifting the resonance wavelengths (Fig.~\ref{fig:sens}). In
each experiment the temperature was decreased to 30$^\circ$C to
estimate the hysteresis. The hysteresis was 1.4, 0.4, 0.05, -1.0 nm,
for $\Lambda_{\rm G} =$ 580, 620, 740, 820 $ \mu$m, respectively.
%---------------------------------------------------------------------------
%---------------------------------------------------------------------------
\begin{center} \begin{table} \begin{tabular}{|c|c|c|c|c|c|c|} \hline
& \bf Air& \multicolumn{4}{|c|}{\bf Methanol} \\
     \cline{2-6} $\Lambda_{\rm G}$ & $\lambda_{\rm r}$ &
$\lambda_{\rm r}$ & $\frac{\ud \lambda_{\rm r}}{\ud n_{\rm
r}}$ & $n_{\rm r}$ & $\frac{\ud n_{\rm r}}{\ud T}$\\
\hline 820 $\mu$m & 623 nm & 671  nm  &415nm&1.3228&-4.067$\times 10^{-4}$\\
\hline 740 $\mu$m & 704 nm & 776  nm  &651nm&1.3157&-3.977$\times 10^{-4}$\\
\hline 620 $\mu$m & 832 nm & 929  nm  &1140nm&1.3093&-3.896$\times 10^{-4}$\\
\hline 580 $\mu$m & 923 nm & 1050 nm  &1460nm&1.3061 & -3.856$\times 10^{-4}$\\
\hline \end{tabular} \caption{Resonant wavelengths, $\lambda_{\rm
r}$, and grating periods, $\Lambda_{\rm G}$. The resonant
wavelengths for methanol filled PCF-LPGs and the corresponding
material parameters $n_{\rm r}$ and $\frac{\ud \lambda_{\rm r}}{\ud
n_{\rm r}}$  for methanol at the resonance
wavelength.}\label{table}\end{table} \end{center}
%---------------------------------------------------------------------------
%---------------------------------------------------------------------------
%---------------------------------------------------------------------------
%---------------------------------------------------------------------------
 The sensitivity
increases more than three times from the resonance wavelength 650 nm
to 1050 nm, as seen in the inset of Fig.~\ref{fig:sens}. We will
allude to this dramatic increase in the following.
%---------------------------------------------------------------------------
\begin{figure}[b!] \begin{center} \includegraphics[angle = 0, width
=0.5\textwidth]{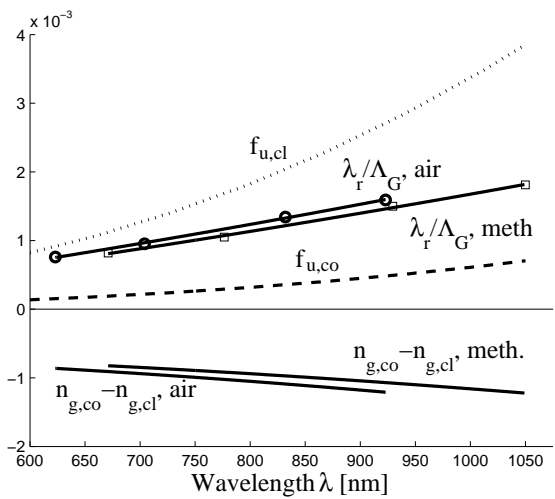} \end{center} \caption{The group
index mismatch calculated from the resonance wavelengths and grating
periods for air and methanol filled PCF-LPG. The field fraction as
function of wavelength for the core and cladding
mode.}\label{fig:ng} \end{figure}
%---------------------------------------------------------------------------
 It may also seem surprising that the resonance wavelength is
redshifted rather than blueshifted when the methanol is infiltrated
into the PCF-LPG. One would suspect that the cladding index was
increased more than the core index, and that the resonance
wavelength should be blue shifted according to the resonance
condition. The supposed anomaly originates in the wavelength
dependence of the effective indices. The wavelength dependence
causes the resonance wavelength to decrease with increasing grating
period as seen in Table \ref{table}. A consistent treatment of the
resonance condition with the chain rule \cite{shu2002} yields the
surprising result for the resonant wavelength shift
%---------------------------------------------------------------------------
$\frac{\ud \lambda_{\rm r}}{\ud n_{\rm r}} = \frac {\lambda_{\rm r}}
{n_{\rm g,co}- n_{\rm g,cl}} \frac{\ud (n_{\rm co}-n_{\rm cl})}{\ud
n_{\rm r}}$
%---------------------------------------------------------------------------
, where $n_{\rm r}$ is the refractive index of methanol and $n_{{\rm
g},i} = n_{i} -\lambda \partial_\lambda n_{i}$ is the group index of
mode $i$. The group index mismatch, $n_{\rm g,co}- n_{\rm g,cl}$,
can be calculated from the resonance wavelengths and the grating
period by $n_{\rm g,co}- n_{\rm g,cl} = \lambda_{\rm r}/\Lambda_{\rm
G} - \lambda_{\rm r} \partial_{\lambda}( \lambda_{\rm
r}/\Lambda_{\rm G}) $, to obtain the curves in Fig.~\ref{fig:ng}.
Clearly, the group index mismatch is negative, and this accounts for
the redshifting of the resonance wavelength for increasing
refractive index, since $ \frac{\ud (n_{\rm co}-n_{\rm cl})}{\ud
n_{\rm r}}$ is negative giving an overall positive sign. Using
perturbation theory it is possible to derive an analytical
expression for the sensitivity \cite{rindorf2006_8}
%---------------------------------------------------------------------------
 \begin{eqnarray}\frac{\ud \lambda_{\rm r}}{\ud
n_{\rm r}} &\propto& \frac{\ud (n_{\rm co}-n_{\rm cl})}{\ud n_{\rm
r}} \simeq \frac {2n_{\rm co}}{n_{\rm r}}(f_{\rm u,co}- f_{\rm
u,cl}), \end{eqnarray}
%---------------------------------------------------------------------------
 where $f_{{\rm u},i}$ is the fraction of
field inside the holes of the PCF of mode $i$. The fraction for the
core and cladding mode has been calculated using a commercial
implementation of the finite element method \cite{comsol}. The
fraction increases, as expected, with wavelength. The cladding
fraction is much larger than the core fraction, Fig.~\ref{fig:ng}.
Thus the contribution to the wavelength shift exclusively determined
by the perturbation the cladding mode.
%---------------------------------------------------------------------------

%---------------------------------------------------------------------------
It is enticing  to design a LPG-PCF with as a high sensitivity by
adjusting the PCF structure. One such opportunity is obtaining group
index matching, $n_{\rm g,co}- n_{\rm g,cl} \sim 0$, since the
resonance wavelength shift is inversely proportional to this as
mentioned earlier. Such a sensor could thus have extremely high
sensitivity. Unfortunately, the full-width half maximum of the
resonance dip also depends on the group index mismatch, and such a
PCF-LPG will have very wide resonance dips \cite{rindorf2006_8}.
Taking this fact into account it is still possible to enhance the
sensitivity by designing the PCF structure.
%---------------------------------------------------------------------------

%---------------------------------------------------------------------------
The optimized PCF used by Phan Huy \etal\ \cite{phanhuy2007}
achieved an increase in sensitivity by two orders of magnitude with
respect to a large core design. We anticipate that a similar
increase could be realized for a PCF-LPG. Indeed, theoretical
considerations predict a minimal detectable refractive index change
of 10(-7) RIU \cite{rindorf2006_8}, comparable to the best surface
plasmon resonance biosensors. This could lead to competitive
PCF-LPGs biosensors \cite{rindorf2006_7}.

%--------------------------------------------------------------------
% ------------ Conclusion ------------
%--------------------------------------------------------------------
In conclusion, we have demonstrated highly sensitive refractometers
in photonic crystal fiber with a long-period grating by infiltrating
the fluid into to the fiber. The sensitivity is increased by a
factor of three from the resonance wavelengths 671 to 1050 nm.
%---------------------------------------------------------------------------

%--------------------------------------------------------------------
%\section*{Acknowledgments}

We acknowledge Crystal-Fibre A/S \cite{cf} for providing the PCF.
%--------------------------------------------------------------------
%\bibliographystyle{osajnl}
\bibliographystyle{apsrev}

\begin{thebibliography}{12}
\expandafter\ifx\csname natexlab\endcsname\relax\def\natexlab#1{#1}\fi
\expandafter\ifx\csname bibnamefont\endcsname\relax
  \def\bibnamefont#1{#1}\fi
\expandafter\ifx\csname bibfnamefont\endcsname\relax
  \def\bibfnamefont#1{#1}\fi
\expandafter\ifx\csname citenamefont\endcsname\relax
  \def\citenamefont#1{#1}\fi
\expandafter\ifx\csname url\endcsname\relax
  \def\url#1{\texttt{#1}}\fi
\expandafter\ifx\csname urlprefix\endcsname\relax\def\urlprefix{URL }\fi
\providecommand{\bibinfo}[2]{#2}
\providecommand{\eprint}[2][]{\url{#2}}

\bibitem[{\citenamefont{Bhatia and Vengsarkar}(1996)}]{bhatia1996}
\bibinfo{author}{\bibfnamefont{V.}~\bibnamefont{Bhatia}} \bibnamefont{and}
  \bibinfo{author}{\bibfnamefont{A.~M.} \bibnamefont{Vengsarkar}},
  \bibinfo{journal}{Opt. Lett.} \textbf{\bibinfo{volume}{21}},
  \bibinfo{pages}{692} (\bibinfo{year}{1996}).

\bibitem[{\citenamefont{Chong et~al.}(2004)\citenamefont{Chong, Shum, Haryono,
  Yohana, Rao, Lu, and Zhu}}]{zhu2004b}
\bibinfo{author}{\bibfnamefont{J.~H.} \bibnamefont{Chong}},
  \bibinfo{author}{\bibfnamefont{P.}~\bibnamefont{Shum}},
  \bibinfo{author}{\bibfnamefont{H.}~\bibnamefont{Haryono}},
  \bibinfo{author}{\bibfnamefont{A.}~\bibnamefont{Yohana}},
  \bibinfo{author}{\bibfnamefont{M.~K.} \bibnamefont{Rao}},
  \bibinfo{author}{\bibfnamefont{C.}~\bibnamefont{Lu}}, \bibnamefont{and}
  \bibinfo{author}{\bibfnamefont{Y.}~\bibnamefont{Zhu}}, \bibinfo{journal}{Opt.
  Com.} \textbf{\bibinfo{volume}{229}}, \bibinfo{pages}{65}
  (\bibinfo{year}{2004}).

\bibitem[{\citenamefont{Eggleton et~al.}(1999)\citenamefont{Eggleton,
  Westbrook, Windeler, Sp\"alter, and Strasser}}]{eggleton:1999}
\bibinfo{author}{\bibfnamefont{B.~J.} \bibnamefont{Eggleton}},
  \bibinfo{author}{\bibfnamefont{P.~S.} \bibnamefont{Westbrook}},
  \bibinfo{author}{\bibfnamefont{R.~S.} \bibnamefont{Windeler}},
  \bibinfo{author}{\bibfnamefont{S.}~\bibnamefont{Sp\"alter}},
  \bibnamefont{and} \bibinfo{author}{\bibfnamefont{T.~A.}
  \bibnamefont{Strasser}}, \bibinfo{journal}{Opt. Lett.}
  \textbf{\bibinfo{volume}{24}}, \bibinfo{pages}{1460 } (\bibinfo{year}{1999}).

\bibitem[{\citenamefont{Fini}(2004)}]{fini2004}
\bibinfo{author}{\bibfnamefont{J.~M.} \bibnamefont{Fini}},
  \bibinfo{journal}{Meas. Sci. \& Technol.} \textbf{\bibinfo{volume}{15}},
  \bibinfo{pages}{1120} (\bibinfo{year}{2004}).

\bibitem[{\citenamefont{Yariv}(1973)}]{yariv1973}
\bibinfo{author}{\bibfnamefont{A.}~\bibnamefont{Yariv}}, \bibinfo{journal}{IEEE
  J. Quant. Electron.} \textbf{\bibinfo{volume}{QE-9}}, \bibinfo{pages}{919 }
  (\bibinfo{year}{1973}).

\bibitem[{\citenamefont{Zhu et~al.}(2003)\citenamefont{Zhu, Shum, Chong, Rao,
  and Lu}}]{zhu2003}
\bibinfo{author}{\bibfnamefont{Y.}~\bibnamefont{Zhu}},
  \bibinfo{author}{\bibfnamefont{P.}~\bibnamefont{Shum}},
  \bibinfo{author}{\bibfnamefont{J.-H.} \bibnamefont{Chong}},
  \bibinfo{author}{\bibfnamefont{M.~K.} \bibnamefont{Rao}}, \bibnamefont{and}
  \bibinfo{author}{\bibfnamefont{C.}~\bibnamefont{Lu}}, \bibinfo{journal}{Opt.
  Lett.} \textbf{\bibinfo{volume}{28}}, \bibinfo{pages}{2467}
  (\bibinfo{year}{2003}).

\bibitem[{\citenamefont{El-Kashef}(2000)}]{el-kashef}
\bibinfo{author}{\bibfnamefont{H.}~\bibnamefont{El-Kashef}},
  \bibinfo{journal}{Physica B} \textbf{\bibinfo{volume}{279}},
  \bibinfo{pages}{295} (\bibinfo{year}{2000}).

\bibitem[{\citenamefont{Shu et~al.}(2002)\citenamefont{Shu, Zhang, and
  Bennion}}]{shu2002}
\bibinfo{author}{\bibfnamefont{X.~W.} \bibnamefont{Shu}},
  \bibinfo{author}{\bibfnamefont{L.}~\bibnamefont{Zhang}}, \bibnamefont{and}
  \bibinfo{author}{\bibfnamefont{I.}~\bibnamefont{Bennion}},
  \bibinfo{journal}{J. Lightwave Technol.} \textbf{\bibinfo{volume}{20}},
  \bibinfo{pages}{255} (\bibinfo{year}{2002}).

\bibitem[{\citenamefont{Rindorf and Bang}()}]{rindorf2006_8}
\bibinfo{author}{\bibfnamefont{L.}~\bibnamefont{Rindorf}} \bibnamefont{and}
  \bibinfo{author}{\bibfnamefont{O.}~\bibnamefont{Bang}},
  \bibinfo{journal}{Submitted}  (????).

\bibitem[{com()}]{comsol}
\bibinfo{note}{Http://www.comsol.com}.

\bibitem[{\citenamefont{Huy et~al.}(2007)\citenamefont{Huy, Laffont,
  Dewynter-Marty, Ferdinand, Roy, Auguste, Pagnoux, Blanc, and
  Dussardier}}]{phanhuy2007}
\bibinfo{author}{\bibfnamefont{M.~C.~P.} \bibnamefont{Huy}},
  \bibinfo{author}{\bibfnamefont{G.}~\bibnamefont{Laffont}},
  \bibinfo{author}{\bibfnamefont{V.}~\bibnamefont{Dewynter-Marty}},
  \bibinfo{author}{\bibfnamefont{P.}~\bibnamefont{Ferdinand}},
  \bibinfo{author}{\bibfnamefont{P.}~\bibnamefont{Roy}},
  \bibinfo{author}{\bibfnamefont{J.-L.} \bibnamefont{Auguste}},
  \bibinfo{author}{\bibfnamefont{D.}~\bibnamefont{Pagnoux}},
  \bibinfo{author}{\bibfnamefont{W.}~\bibnamefont{Blanc}}, \bibnamefont{and}
  \bibinfo{author}{\bibfnamefont{B.}~\bibnamefont{Dussardier}},
  \bibinfo{journal}{Opt. Lett.} \textbf{\bibinfo{volume}{32}},
  \bibinfo{pages}{2390} (\bibinfo{year}{2007}).

\bibitem[{\citenamefont{Rindorf et~al.}(2006)\citenamefont{Rindorf, Jensen,
  Dufva, Pedersen, Hoiby, and Bang}}]{rindorf2006_7}
\bibinfo{author}{\bibfnamefont{L.}~\bibnamefont{Rindorf}},
  \bibinfo{author}{\bibfnamefont{J.~B.} \bibnamefont{Jensen}},
  \bibinfo{author}{\bibfnamefont{M.}~\bibnamefont{Dufva}},
  \bibinfo{author}{\bibfnamefont{L.~H.} \bibnamefont{Pedersen}},
  \bibinfo{author}{\bibfnamefont{P.~E.} \bibnamefont{Hoiby}}, \bibnamefont{and}
  \bibinfo{author}{\bibfnamefont{O.}~\bibnamefont{Bang}},
  \bibinfo{journal}{Opt. Express} \textbf{\bibinfo{volume}{14}},
  \bibinfo{pages}{8824} (\bibinfo{year}{2006}).

\end{thebibliography}

\end{document}